\documentstyle[12pt,psfig]{article}
\normalbaselines
\begin{document}

\newcommand{\sign}{\, \textrm{sign} \,}

\vbox {\vspace{6mm}}

\begin{center}
{\large \bf NONCLASSICAL EVOLUTION OF A FREE PARTICLE}\\[7mm]

Lars M. Johansen \footnote{Permanent address: Buskerud College,
P.O. Box 251, N-3601 Kongsberg, Norway} \\
{\it
Institute of Physics, University of Oslo, \\
P.O.Box 1048 Blindern, N-0316 Oslo, Norway
}\\[5mm]
\end{center}

\vspace{2mm}

\begin{abstract}

A conditional kinetic energy is defined in terms of the Wigner
distribution. It is shown that this kinetic energy may become
negative for negative Wigner distributions. A free particle wave
packet with negative kinetic energy will spread in a nonclassical
manner.

\end{abstract}

\section{The Free Particle}

In classical statistical mechanics, a particle is described in terms
of a nonnegative phase space distribution
\begin{equation}
	P(x,p;t) \ge 0.
	\label{eq:posdist}
\end{equation}
This is the probability of the particle having position $x$ and
momentum $p$ at the time $t$. The time evolution of the phase space
distribution for a free particle is
\begin{equation}
	{\partial P \over \partial t} = - {p \over m} \, {\partial P
	\over \partial x}.
	\label{eq:classtraj}
\end{equation}
That quantum mechanics can be formulated in a similar manner was
first discovered by Wigner \cite{Wigner32}. However, the Wigner
distribution $W(x,p;t)$ has the nonclassical feature of being
negative for certain quantum states. In fact, among pure states only
the gaussian states have nonnegative Wigner distributions
\cite{Hudson74}. However, even if a free particle is in a state with
negative Wigner distribution, it will not necessarily behave
exhibit different \emph{dynamics} than an ensemble of classical free
particles \cite{Lee82}. For instance, the Wigner distribution obeys
the same equation of motion (\ref{eq:classtraj}) as a classical phase
space distribution.

Bracken and Melloy \cite{Bracken94} recently found that the
probability of observing a free particle in a certain region of space
may increase even though there is zero probability for the particle
to have momenta pointing towards this region. They explained this
effect in terms of negative probability.

Here we shall study another dynamical effect which defies explanation
in terms of the classical model (\ref{eq:posdist}) and (\ref{eq:classtraj}). To this end, we consider the modulus of $x$,
\begin{equation}
	\langle \mid x \mid \rangle = \int dp \int dx \,
	| \, x \, | \, W(x,p;t).
\end{equation}
We now find that
 \begin{equation}
 	{d \langle \mid x \mid \rangle \over dt} = {1 \over m} \int dp \,
 	p \int dx \, \sign x \, W(x,p;t),
 \end{equation}
and
\begin{equation}
	 	{d^2 \langle \mid x \mid \rangle \over dt^2} = {2 \over m^2}
	 	\,\int dp \, p^2 \, W(0,p;t).
		\label{eq:secder}
\end{equation}
For an ensemble of classical, free particles we have the condition
\begin{equation}
	{d^2 \langle \mid x \mid \rangle \over dt^2} \ge 0,
	\label{eq:nonneg}
\end{equation}
since the phase space distribution must be nonnegative. $\langle | x
| \rangle$ is a measure of the \emph{uncertainty} in position
provided that $\langle x \rangle=0$. It is often called the absolute
deviation. According to Eq. (\ref{eq:nonneg}), the curvature with
respect to time of the absolute deviation must be nonnegative in
classical theory. This inequality therefore sets a constraint on the
dynamics of the spreading of a wave packet in classical mechanics.

What is the physical significance of the r.h.s. of Eq.
(\ref{eq:secder})? We introduce the moments $\pi_n$ defined by
\begin{equation}
	\pi_n(x;t) = \int dp \, p^n \, W(x,p;t).
	\label{eq:kindens}
\end{equation}
$\pi_n(x;t)/\pi_0(x;t)$ can be interpreted classically as the average
of $p^n$ given $x$ \cite{Moyal49}. Violation of (\ref{eq:nonneg})
therefore can be interpreted in classical terms as due to negative
kinetic energy given $x$.

It may seem surprising that a negative kinetic energy can be
observed. Indeed, the operator $\hat{p}^2$ has only nonnegative
eigenvalues, and the expectation of this operator therefore is always
nonnegative. But this does not imply that also \emph{conditional}
kinetic energy must be nonnegative. Indeed, in tunneling, it seems
as if the particle traversing the tunneling region has negative
kinetic energy, since the total energy is lower than the energy of
the potential barrier. But also here one always finds a nonnegative
kinetic energy. However, Aharonov \emph{et al.} \cite{Aharonov93}
have shown that if the kinetic energy is first measured followed by a
position measurement, the subensemble of particles found in the
tunneling region may display negative kinetic energy.

Finally, let's study a system with negative kinetic energy. To this
end, consider the (unnormalized) state
\begin{equation}
	| \psi \rangle = | \alpha \rangle + | - \alpha \rangle.
\end{equation}
This is a superposition of two coherent states $180^{\circ}$ out of
phase with respect to each other. With a choice of units so that
$\hbar=1$, it has a Wigner distribution
\begin{eqnarray}
	W(x,p;0) = {1 \over \pi} \left [ e^{-(p-p_0)^2 -
	(x-x_0)^2} + e^{-(p+p_0)^2 - (x+x_{\alpha})^2} + 2 e^{-x^2 - p^2}
	\cos 2(p_0 x - p x_0) \right ].
	\label{eq:wigner}
\end{eqnarray}
This distribution has negative regions. For the choice
\mbox{$x_0=0$}, these regions are centered along the line
\mbox{$p=0$}, and for \mbox{$p_0=0$} they are centered along
\mbox{$x=0$}. As seen from Eq. (\ref{eq:secder}), violation of
inequality (\ref{eq:nonneg}) for \mbox{$t=0$} requires
that the Wigner distribution has negative regions for \mbox{$x=0$}.
We therefore use the parameter choice \mbox{$p_0=0$}.

\begin{centering}
\centerline{\psfig{figure=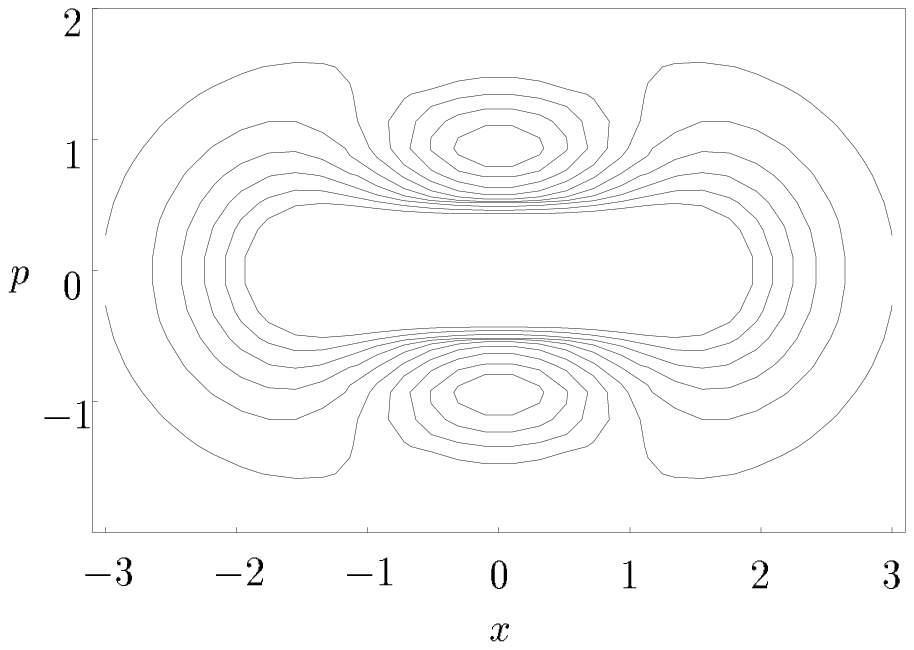,height=6cm}}
\end{centering}

\begin{quotation}
FIG. 1.  Contourplot of the Wigner distribution for an even
coherent state where $x_0=\sqrt{2}$ and $p_0=0$. Note the negative
regions along $x=0$.
\end{quotation}

The solution of Eq. (\ref{eq:classtraj}) is $W(x,p;t)=W(x-pt/m,p;0)$.
Using this and Eq. (\ref{eq:wigner}), we get from Eq.
(\ref{eq:kindens}) by integration
\begin{equation}
	\pi_2(0;t) = {1 \over \sqrt{\pi}} \: \exp \left (-
	{x_0^2 \over 1+t^2} \right ) \: {1 - x_0^2 + t^2 + x_0^2 t^2
	\over (1+t^2)^{5/2}},
\end{equation}
where we have assumed that $p_0=0$. We see that $\pi_2$ is negative
if
\begin{equation}
	t < \sqrt{x_0^2 - 1 \over x_0^2 + 1}.
	\label{eq:timelimit}
\end{equation}
Thus $\pi_2$ may become negative provided that $x_0>1$, and it has a
relative minimum for $t=0$ and $x_0=\sqrt{2}$.

\begin{centering}
\centerline{\psfig{figure=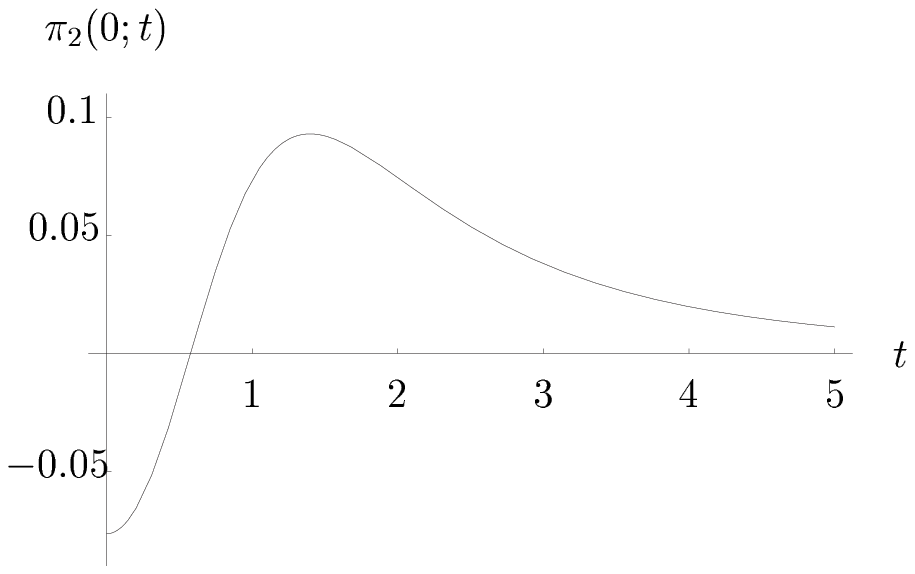,height=6cm}}
\end{centering}

\begin{quotation}
FIG. 2.	 Plot of $\pi_2(0;t)$ for a free particle in an even coherent
state, where $x_0=\sqrt{2}$ and $p_0=0$. It is negative for $t <
1/\sqrt{3}$. This is impossible for an ensemble of classical, free
particles, since it implies negative kinetic energy. According to Eq.
(\ref{eq:secder}), $\pi_2(0;t)$ is also proportional to the curvature
of the expected absolute value of position.
\end{quotation}

\section{Quantum Optics}

We have found that violation of inequality (\ref{eq:nonneg}) can
only be explained in terms of a negative Wigner distribution. Let's
now consider a simple realization of a similar scheme in quantum
optics.

Consider the rotated quadrature variable
\begin{equation}
	x_{\theta} = x \cos \theta + p \sin \theta.
\end{equation}
This observable can be measured in homodyne detection, if the
radiation mode described by $W(x,p)$ is mixed with a strong local
oscillator with phase $\theta$ \cite{Yuen83b,Schumaker84}. We may
``simulate" free particle evolution by introducing the variable
\cite{Leonhardt95b}
\begin{equation}
	\chi_{\tau} = {x_{\theta} \over \cos \theta} = x + p \tau,
\end{equation}
where
\begin{equation}
	\tau = \tan \theta.
\end{equation}
We thus have
\begin{equation}
	\langle \mid \chi_{\tau} \mid \rangle = \int dp \int dx \mid x +
	p \tau \mid W(x,p).
	\label{eq:ordprob}
\end{equation}
We substitute $x' = x + p \tau$, so that
\begin{equation}
	\langle \mid \chi_{\tau} \mid \rangle = \int dp \int dx' \mid x'
	\mid W(x'-p \tau,p).
\end{equation}
This clarifies that there is no ordering problem associated with Eq.
(\ref{eq:ordprob}) \cite{Moyal49}. We may now proceed to demonstrate
that
\begin{equation}
	{d^2 \langle \mid \chi_{\tau} \mid \rangle \over dt^2} = 2 \int
	dp \; p^2 \; W(0,p).
\end{equation}
In analogy with the free particle case, we therefore see that
\begin{equation}
	{d^2 \langle \mid \chi_{\tau} \mid \rangle \over d \tau^2}
	\ge 0 \label{eq:quadrature}
\end{equation}
for nonnegative Wigner distributions. Violation of this inequality
indicates that the Wigner distribution has negative regions along
$x=0$. It can be tested in homodyne detection.

\section*{Conclusion}

We have seen that it is possible to observe negative kinetic energy
for free particles. It was shown that this leads to nonclassical
evolution of the position absolute deviation. The scheme was also
applied to quantum optics, where a simple experiment was proposed to
detect negative Wigner distributions.

Depending on the operator ordering we assign to a scalar product of
$x$ and $p$, we get a different quasi phase space distribution
\cite{Cahill69}. Thus, since there is no unique phase space
distribution in quantum mechanics, there is no unique conditional
kinetic energy either. It turns out, e.g., that the state examined by
Aharonov {\em et al.} \cite{Aharonov93} does not give a negative
Wigner kinetic energy. An essential part of the analysis of Aharonov
{\em et al.} was an inherent uncertainty in the pointer position of
their measurement apparatus.

When the classical model expressed by Eqs. (\ref{eq:posdist}) and
(\ref{eq:classtraj}) breaks down, one may abandon either assumption
(\ref{eq:posdist}) or (\ref{eq:classtraj}). By using the Wigner
distribution, we abandon (\ref{eq:posdist}) and the concept of
nonnegative probability. Using other distributions, one might instead
abandon (\ref{eq:classtraj}) \cite{Lee82}. This amounts to
abandoning the idea that a point in phase space moves with constant
velocity.

The analysis done here can be generalized to particles in arbitrary
potentials. Also in this case it can be shown that negative kinetic
energy leads to nonclassical evolution of the position absolute value.

\section*{Acknowledgments}

I would like to thank Howard M. Wiseman and Stefan Weigert for
drawing my attention to the paper by Bracken and Melloy
\cite{Bracken94}. I would also like to thank Ulf Leonhardt for useful
comments.

\end{document}